\documentclass[multphys,vecphys]{svmult}

\usepackage{makeidx}         
\usepackage{graphicx}        
\usepackage{multicol}        
\makeindex

\begin{document}

\title*{Heat conduction in one dimensional systems: Fourier
law, chaos, and heat control}
\author{Giulio Casati\inst{1,2}\and
Baowen Li\inst{2}}

\institute{ Center for Nonlinear and Complex Systems,
Universita' degli studi dell'Insubria, Como\\
Istituto Nazionale di Fisica della Materia, Unita' di Como, and \\
Istituto Nazionale di Fisica Nucleare, sezione di Milano, Milano,
Italy\\
\texttt{Giulio.Casati@uninsubria.it, phycg@nus.edu.sg} \and
Department of Physics, National University of Singapore, Singapore 117542, Republic of Singapore\\
\texttt{phylibw@nus.edu.sg}
}

\maketitle

\section{Introduction}

 ``{\it It seems there is no problem in modern
physics for which there are on record as many false starts, and as
many theories which overlook some essential feature, as in the
problem of the thermal conductivity of nonconducting crystals}"
\cite{Peierls}. This statement by R. Peierls goes back to almost
50 years ago, yet it appears to be still valid. Compared with
charge flow (electric current), much less is known about the heat
flow.

Nevertheless, encouraging results have been obtained in recent
years (see reviews\cite{Review}  and the references therein). For
example it is now known that in one dimensional systems of the
Fermi-Pasta-Ulam (FPU) type \cite{FPU}, heat conduction is
anomalous and the coefficient of thermal conductivity $\kappa$
diverges with the system size $L$ as $\kappa \sim L^{2/5}$ (when
the transverse motion is considered $\kappa \sim L^{1/3}$
\cite{WangLi04}). The connection between anomalous conductivity
and anomalous diffusion has been also
established\cite{LW03,LiChaos05}, which implies in particular that
a subdiffusive system is an insulator in the thermodynamic limit
and a ballistic system is a perfect thermal conductor, the Fourier
law being therefore valid only when phonons undergo a normal
diffusive motion. More profoundly, it has been clarified that
exponential dynamical instability is a
sufficient\cite{Casati84,Casati99} but not a necessary condition
for the validity of Fourier law\cite{LWH02,Alonso02,LCW03,LCWP04}.
These basic studies not only enrich our knowledge of the
fundamental transport laws in statistical mechanics, but also open
the way for applications such as designing novel thermal materials
and/or devices such as the thermal rectifier\cite{rectifier,diode}
and the thermal transistor\cite{transt}.

In this paper we give a brief review of the relation between
microscopic dynamical properties and the Fourier law of heat
conduction as well as  the connection between anomalous conduction
and anomalous diffusion. We then discuss the possibility to
control the heat flow.

\section{Microscopic dynamics and the Fourier heat law}

In spite of several years of investigation, so far there is
neither phenomenological nor fundamental transport theory which
can predict whether or not a given classical, many-body
Hamiltonian system, yields an energy transport governed by the
Fourier heat law. It is known that heat flow obeys a simple
diffusion equation which can be regarded as the continuum limit of
a discrete random walk. However, modern ergodic theory tells us
that for K-systems, a sequence of measurements with finite
precision mimics a truly random sequence and therefore these
systems appear precisely those deterministically random systems
tacitly required by transport theory. It is therefore interesting
to establish which class, if any, of many-body systems satisfy the
necessary stringent requirements for the validity of Fourier law
of heat conduction.

\subsection{Dynamical chaos is a sufficient condition for the Fourier law}

\subsubsection{Ding-a-ling model}

The first example for which convincing evidence has been provided
that Fourier law can be derived on purely dynamical grounds,
without any additional statistical assumptions, is the so-called
ding-a-ling model proposed in \cite{Casati84}.

The Hamiltonian of the ding-a-ling model is:

\begin{equation}
H=\frac{1}{2}\sum \left(p^2_k+ \omega^2_kq_k^2\right) + \mbox{hard} \hspace{.2cm}
\mbox{point}\hspace{.2cm} \mbox{core}
\label{eq:Ding-a-Ling}
\end{equation}

where $\omega_k $ equals $\omega$ for even $k$ and zero for odd
$k$ and where all particles have unit mass. It can be shown that
it is possible to fix the energy per particle ($E=1$) and the half
of the lattice distance between two bound particles ($l_0=1$).
After this the dynamics is uniquely determined by the frequency
$\omega$ which becomes therefore the only adjustable parameter.

\begin{figure}
\begin{center}
\includegraphics[width=7cm]{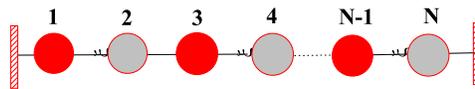}
\vspace{-3cm}
 \caption {The N-particle ding-a-ling model.
Odd particles can move freely in between two collisions, while
even particles are bounded by a harmonic spring.}
\end{center}
\end{figure}

 As it seen in Fig 1. the model is a one-dimensional array
of equal mass, hard-point particles, the even-numbered particles
form a set of equally spaced harmonic oscillators with each
oscillator bound to its individual lattice site and with all
oscillators vibrating with the same frequency $\omega$. The
odd-numbered particles are free particles constrained only by the
two adjacent even-numbered oscillators.

This model has been chosen in order to meet two requirements.
First, it is necessary to select a deterministically random
system, and second, the system must be sufficiently simple to
allow efficient numerical analysis. Let us also recall that even
systems obeying the Fourier heat law can transport energy in the
form of slowly decaying coherent excitations such as soundlike
pulses. In numerical experiments, which unavoidably consider only
a small number of particles, this phenomenon is quite troublesome.
Though K or almost K-system guarantees that these soundlike
solutions will eventually decay, one needs to find a small chaotic
system in which this decay rate is sufficiently rapid\cite{boris}.
The selected model (1) is a many-body system which exhibits, as
the frequency is varied, the full range of behaviour from
integrable to almost K and which at the same time has no problem
with energy-bearing, long-lived, solitonlike pulses. Indeed, it
can be shown that as $\omega$ is increased from zero (which is the
well known integrable 1-d hard point gas), the system undergoes a
transition from integrable to almost fully chaotic. This fact
makes this model ideal to study the connection between Fourier law
and microscopic dynamical chaos.

Heat conductivity has been studied by placing the end particles in
contact with two thermal reservoirs at different temperatures (see
\cite{Casati84} for details)and then integrating the equations of
motion. Numerical results \cite{Casati84} demonstrated that, in
the small $\omega$ regime, the heat conductivity is system size
dependent, while at large $\omega$, when the system becomes almost
fully chaotic, the heat conductivity becomes independent of the
system size (if the size is large enough). This means that Fourier
law is obeyed in the chaotic regime.

\subsubsection{Lorentz gas channel}

The above conclusions have been nicely confirmed and clarified in
\cite{Casati99} where the heat conduction has been studied in a
Lorentz gas channel - a quasi one dimensional billiard with
circular scatterers. The model (see Fig. 2) consists of two
parallel lines and a series of semicircles of radius R placed in a
triangular lattice along the channel. By construction no particle
can move along the horizontal direction without colliding with the
disks. The dynamics in the Lorentz gas is rigorously known to be
mixing and all trajectories with nonzero projection on the
horizontal direction are of hyperbolic type; further it has
positive Kolmogorov-Sinai entropy and a well defined diffusive
behavior. Very accurate numerical evidence has been provided
\cite{Casati99} which shows that heat conduction in this model
obeys the Fourier law. This result therefore clearly indicates
that mixing with positive Lyapounov exponents is a sufficient
condition to ensure Fourier law of heat conduction.

\begin{figure}
\begin{center}
\includegraphics[width=7cm]{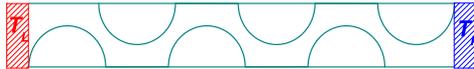}
\vspace{-1.5cm}
 \caption {The geometry of the Lorentz gas channel
model. The two heat reservoirs at temperatures $T_L$ and $T_R$ are
indicated.}
\end{center}
\end{figure}

\subsection{Chaos is not a necessary condition}

\subsubsection{Triangle billiard gas channel}

Quite naturally, the next question which arises is whether strong,
exponential unstable chaos, being sufficient, is also necessary.

In this connection let us remark that in spite of several efforts,
the relation between Lyapounov exponents, correlations decay,
diffusive and transport properties is still not completely clear.
For example a model has been presented \cite{triangle1} which has
zero Lyapounov exponent and yet it exhibits unbounded Gaussian
diffusive behavior. Since diffusive behavior is at the root of
normal heat transport then the above result\cite{triangle1}
constitutes a strong suggestion that normal heat conduction can
take place even without the strong requirement of exponential
instability.

\begin{figure}
\begin{center}
\includegraphics[width=7cm]{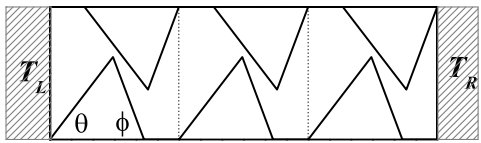}
\caption {The geometry of the triangle billiard gas channel.
Particles move in the region outside the triangular scatterers.
The two heat reservoirs at temperatures $T_L$ and $T_R$ are
indicated.}
\end{center}
\end{figure}

The above problem has been addressed in \cite{LCW03}, where we
have considered a quasi-one dimensional billiard model which
consists of two parallel lines and a series of triangular
scatterers (see Fig.3). In this geometry, no particle can move
between the two reservoirs without suffering elastic collisions
with the triangles. Therefore this model is analogous to the
previous Lorentz gas channel with triangles instead of discs: the
essential difference is that in the triangular model the dynamical
instability is linear and therefore the Lyapounov exponent is
zero.

Strong numerical evidence has been given \cite{casati} that the
motion inside a triangular billiard, with all angles irrational
with $\pi$ is mixing, without any time scale. It is therefore
reasonable to expect that the motion inside the irrational
polygonal area of Fig 3 is diffusive thus leading to normal
conductivity.

Indeed, numerical results in \cite{LCW03} show that in the
irrational case (when the ratio $\theta/\pi$ and $\phi/\pi$ are
irrational numbers) the system in Fig 3 exhibits normal diffusion
and the heat conduction obeys the Fourier law. In the rational
case instead, the system shows a superdiffusive behavior, $\langle
\sigma ^2\rangle=2Dt^{1.178}$ \cite{LCW03}and the heat
conductivity diverges with the system size as $\kappa\approx
N^{0.25\pm 0.01}$.

\subsubsection{Alternate mass-core hard potential channel}

In the two billiard gas models just discussed there is no local
thermal equilibrium. Even though the internal temperature can be
clearly defined at any position\cite{Casati99}, the above property
may be considered unsatisfactory\cite{Dhar299}. In order to
overcome this problem, we have recently introduced a similar model
which however exhibits local thermal equilibrium, normal
diffusion, and zero Lyapunov exponent\cite{LCWP04}.

\begin{figure}
\begin{center}
\includegraphics[width=7cm]{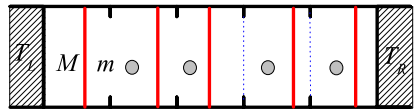}
\caption {The geometry of the alternate mass-core hard potential
channel. The elementary cell is indicated by the two dotted lines.
The bars have mass $M=1$, and the particles have mass $m=
(\sqrt{5}-1)/2$. The two heat baths at temperatures $T_L$ and
$T_R$ are indicated. }
\end{center}
\end{figure}
This model consists of a one-dimensional chain of elastically
colliding particles with alternate masses $m$ and $M$. In order to
prevent total momentum conservation we confine the motion of
particles of mass $M$ (bars) inside separate cells. Schematically
the model is shown in Fig.4: particles with mass $m$ move
horizontally and collide with bars of mass $M$ which, besides
suffering collisions with the particles, are elastically reflected
back at the edges of their cells. In between collisions, particles
and bars move freely.

Numerical results\cite{LCWP04} clearly indicate that this model
also obeys the Fourier law.

\section{Anomalous Heat Conduction}

Numerical experiments have shown that in many one dimensional
systems with total momentum conservation, the heat conduction does
not obey the Fourier law and the heat conductivity depends on the
system size. For example, in the so-called FPU model, $\kappa\sim
L^{\beta}$, with $\beta=2/5$,  and if the transverse motion is
introduced, $\beta=1/3$. Moreover, in the billiard gas channels
(with conserved total momentum), the value of $\beta$ differs from
model to model\cite{LW03}.  The question is whether one can relate
$\beta$ to the dynamical and statistical properties of the system.

\begin{figure}
\begin{center}
\includegraphics[width=7cm]{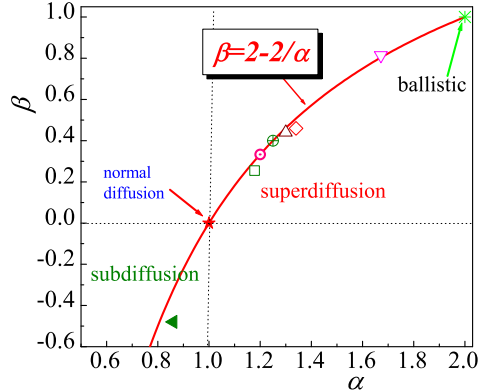}
\vspace{-0.5cm} \caption{ Comparison of prediction (4) with
numerical data. Normal diffusion ($\star$).   The ballistic motion
($\ast$). Superdiffusion: 1D Ehrenfest gas
channel\cite{LWH02}($\bigtriangledown$); the rational triangle
channel\cite{LCW03} (empty box);  the polygonal billiard channel
with $\phi_1=(\sqrt{5}-1)\pi/4$), and $\phi_2=\pi/3$
\cite{Alonso02}($\triangle$); the triangle-square channel
gas\cite{LiChaos05} ($\diamond$). $\beta$ values are obtained from
system size $L\in [192,384]$ for all channels except Ehrenfest
channel\cite{LWH02}. The FPU lattice model at high temperature
regime\cite{LiChaos05}($\odot$), and the single walled nanotubes
at room temperature ($\oplus$).  Subdiffusion: model from
Ref.\cite{Alonso02} (solid left triangle). The solid curve is
$\beta=2-2/\alpha$. }
\end{center}
\end{figure}
Recently, a simple formula has been found \cite{LW03} which
connects anomalous heat conductivity with anomalous diffusion.
More precisely, it has been shown that for a one dimensional
system, if the energy diffusion can be described by
\begin{equation}
\langle \sigma^2\rangle = 2Dt^{\alpha}, \quad(0<\alpha\le 2)
\label{eq:Diffusion}
\end{equation}
then the heat conductivity is given by
\begin{equation}
\kappa =-j/\nabla T \propto L^{\beta}
\label{kappa}
\end{equation}
where the exponent $\beta$ is
\begin{equation}
\beta=2-2/\alpha.
\label{eq:alpha-beta}
\end{equation}

This relation connects heat conduction and diffusion,
quantitatively. As expected, normal diffusion ($\alpha=1$)
corresponds to the size-independent ($\beta=0$) heat conduction
obeying the Fourier law. Moreover, a ballistic motion ($\alpha=2$)
implies that
 the thermal conductivity is proportional to the system size $L$, a
superdiffusive behavior($1<\alpha<2$) implies a divergent thermal
conductivity ($\beta>0$), and a subdiffusive behavior ($\alpha<1$)
implies  zero thermal conductivity ($\beta<0$) in the
thermodynamic limit.

The simple relation (4) is in good agreement with existing data
from billiard gas channels to nonlinear lattices, and even single
walled nanotubes\cite{LiChaos05}. This is shown in Fig. 5, where
we compare the theoretical prediction (4) with existing data in
different models.

We should mention here the one dimensional hard point gas model
for which anomalous heat conduction has been found by several
authors \cite{2mass1,2mass2}.  However it seems there is no
agreement on the numerical value of the exponent $\beta$. Indeed
in Ref.\cite{2mass1}, the value $\beta\approx0.25$ has been found,
while the value $\beta\approx 0.33$ is reported in \cite{2mass2}.

\section{Control of heat flow}

While in the previous sections we have discussed the relation
between dynamical chaos and heat conductivity, in the following we
will turn our attention to the possibility to control heat flow.
Actually a model of thermal rectifier has been recently
proposed\cite{rectifier} in which the heat can flow preferentially
in one direction. Although this model is far away from a prototype
realization, it is based on a mechanism of very general nature
and, as such, is suitable of improvement and may eventually lead
to real applications. This problem is discussed in the next
section.

\subsection{Thermal diode}

In a recent paper\cite{diode}, a thermal diode model has been
proposed in which, even though the underlying physical mechanism
is similar to the one in Ref.\cite{rectifier}, there is a new
crucial element which allows to improve the efficiency by more
than two orders of magnitude.

The diode model consists of two segments of nonlinear lattices
coupled together by a harmonic spring with constant strength
$k_{int}$ (see Fig. 6). Each segment is described by the
(dimensionless) Hamiltonian:
\begin{equation}
H =\sum
\frac{p^2_i}{2m}+\frac{1}{2}k(x_i-x_{i+1}-a)^2-\frac{V}{(2\pi)^2}\cos
2\pi x_i. \label{eq:Ham}
\end{equation}

\begin{figure}
\begin{center}
\includegraphics[width=7.cm]{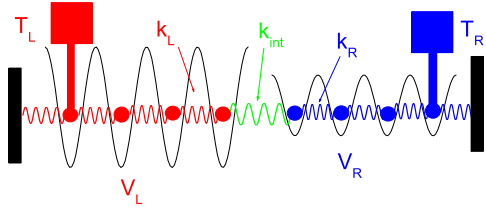}
\vspace{-.5cm} \caption{Configuration of the thermal diode model
based on two coupled FK chains.}
\end{center}
\end{figure}
The two ends of the system are put into contact with thermal baths
at temperature $T_L$ and $T_R$ for left and right  bath,
respectively. In fact, Eq. (\ref{eq:Ham}) is the Hamiltonian of
the Frenkel-Kontorova (FK) model which is known to have normal
heat conduction\cite{HLZ98}. For simplicity we set the mass of the
particles and the lattice constant $m=a=1$. Thus the adjustable
parameters are $(k_L,k_{int},k_R,V_L,V_R,T_L,T_R)$, where the
letter L/R indicates the left/right segment. In order to reduce
the number of adjustable parameters, we set $V_R=\lambda V_L$,
$k_R=\lambda k_L$, $T_L=T_0(1+\Delta), T_R=T_0(1-\Delta)$ and,
unless otherwise stated, we fix  $V_L=5$, $k_L=1$  so that the
adjustable parameters are reduced to four, $(\Delta, \lambda,
k_{int}, T_0)$. Notice that when $\Delta>0$, the left bath is at
higher temperature and vice versa when $\Delta<0$.

 In Fig. 7 we plot the heat current $J$
versus $\Delta$ for different temperatures $T_0$. It is clearly
seen that when $\Delta>0$ the heat current ($J_+$) increases
 with $\Delta$, while in the region  $\Delta<0$  the heat current
 ($J_-$)
 is almost zero, i.e. the system behaves as a  thermal insulator.
 The results in Fig. 7 show that our model has the
  rectifying effect in a wide range of temperatures. The
   rectifying efficiency, defined as $|J_+/J_-|$, could be as high as
   few hundreds times, depending on temperature as well as
   on other parameters.

\begin{figure}
\begin{center}
\includegraphics[width=7.cm]{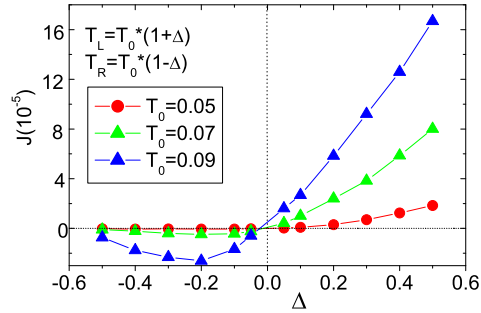}
\vspace{-.5cm}
\caption{Heat current $J$ versus the
dimensionless temperature difference $\Delta$ for different values
of $T_0$. Here the total number of particles $N= 100$,
$k_{int}=0.05$, $\lambda=0.2$. The lines are drawn to guide the
eye.}
\end{center}
\end{figure}

\begin{figure}
\begin{center}
\includegraphics[width=7cm]{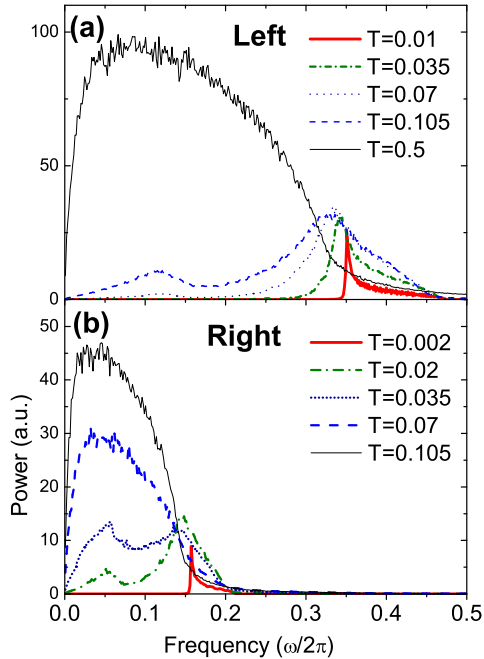}
\caption{Spectra of the two particles at the interface for
different temperatures at $k_{int}=0$. (a) particle at the left
side of the interface, (b) particle at the right side of the
interface. Here $\lambda=0.2$, $N=100$.}
\end{center}
\end{figure}

\subsubsection{Rectifying mechanism}

To understand the underlying rectifying mechanism, let's start
from the energy spectrum of the  interface particles. Fig. 8 shows
the phonon spectra of the left and right interface particles at
different temperature when the two lattices are decoupled
($k_{int}=0$).

The match/mismatch of the energy spectra of the two interface
particles controls  the heat current. It is clearly seen from Fig.
8 that, if the left end is in contact with the high temperature
bath $T_L$, and the right end with the low temperature bath $T_R$
($<T_L$), then the phonon spectra of the two particles at
interface overlap in a large range of frequencies, thus the heat
current can easily go through the system from the left end to the
right end. However, if  the left end is at lower temperature $T_L$
and the right end is  at higher temperature $T_R$ ($>T_L$), then
the phonon spectrum of the right interface particle is mainly in
the low frequency part, while the left interface particle is in
the high frequency part. Then there is almost no overlap in phonon
frequency, and the heat current can hardly go through from right
to left, and the system behaves as an insulator. Why the left and
right particles at the interface have so different phonon spectra?
This can be understood from the following analysis on temperature
dependent phonon spectra, due to the nonlinearity.

(1) {\it Low temperature limit}. At low temperature, the particle
is confined in the valley of the on-site potential.  By
linearizing the equation of motion one can easily obtain the
frequency band\cite{diode}:
\begin{equation}
\sqrt{V}<\omega <\sqrt{V+4k}.
\label{eq:lowfre}
\end{equation}
For  the case of Fig.7 with T=0.01 (left) and T=0.002 (right),
this corresponds to $0.36<\omega/2\pi<0.48$ for the left particle
and to $0.16<\omega/2\pi<0.21$ for the right particle.

As the temperature is increased, the interparticle  potential
$kx^2/2$ becomes more and more important until a critical value
$T_{cr}\approx V/(2\pi)^2$ is reached (we take the Boltzman
constant equal unity), when the kinetic energy is large enough to
overcome the on-site potential barrier. At this point low
frequency appears and this happens at the critical temperatures
$T_{cr}=0.13$ for $V=5$ (left), and $T_{cr}=0.025$ for $V=1$
(right). This is in quite good agreement with the data of Fig.8.

(2) {\it High temperature limit}. In the high temperature limit
the on-site potential can be neglected,  the system is close to
two coupled harmonic chains, and the phonon band is is given
by\cite{kittle}:
\begin{equation}
0<\omega<2\sqrt{k},
\label{eq:highfre}
\end{equation}
which gives $0<\omega/2\pi<0.32$ for the left particle and
$0<\omega/2\pi<0.14$ for the right particle, again in good
agreement with Fig.8.

In fact, in order to optimize the rectifying effect, one should
avoid the overlapping of the phonon bands in the low temperature
limit (Eq.\ref{eq:lowfre}) and that in the high temperature limit
(Eq.\ref{eq:highfre}) for each segment of the system. According to
the above estimates, one should have $V>4k$, which is satisfied
for the case of Fig.8.

We should stress that it is the nonlinearity of the potential that
makes phonon spectra temperature dependent, and thus the
rectifying effect possible. Therefore it is reasonable to expect
the rectifying effect to be present, in different degrees, in
typical nonlinear lattices.

\subsection{Negative Differential Thermal Conductance}

Apart from the ``one-way heat flow", the \textit{negative
differential thermal resistance} phenomenon observed in a certain
temperature intervals in the thermal diode is of particular
interest. As illustrated in Fig.7 for $\Delta<-0.2$, a smaller
temperature difference ($\Delta$), can induce a larger heat
current since, due to nonlinearity, it can result in a better
match in phonon bands.

The same phenomenon is shown for different inter-face coupling
constants $k_{int}$ in Fig. 9(a), and different system size in
Fig. 9(b).

\begin{figure}
\begin{center}
\includegraphics[width=6cm]{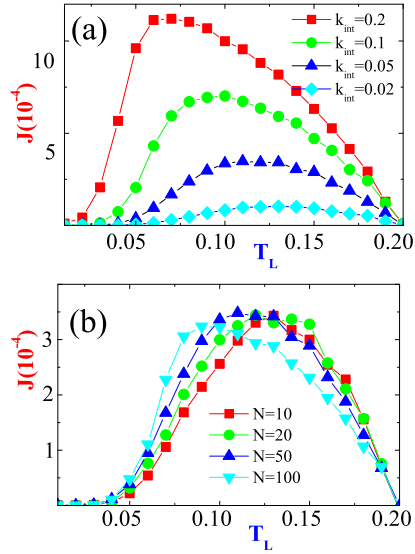}
\caption{  (a) Heat current versus temperature $T_L$ (at fixed
$T_R=0.2$) for different coupling constants, $k_{int}$, with
lattice size $N=50$. The system parameters are: $V_L=5$, $V_R=1$,
$k_L=1$, $k_R=0.2$. (b) Same as (a) but for different system size
$N$. $k_{int}=0.05$. Notice that when $T_L\le 0.1$ the heat
current increases with decreasing the external temperature
difference. }
\end{center}
\end{figure}

\subsection{Thermal Transistor}

The phenomenon of negative differential thermal resistance allows
us to propose a ``thermal transistor"\cite {transt}.
 The configuration of the thermal transistor is shown in Fig.10(a).
It consists of three segments, D, S and G.  The names D, S and G
follow the ones used in a MOSFET (Metal-Oxide-Semiconductor
Field-Effect-Transistor) that is the most important device for
very large scale integrated chips such as microprocessors and
semiconductor memories.  Segment D (from D to O) has a negative
differential thermal resistance in a certain temperature regime
while segment S is a normal heat conductor, i.e., heat current
inside this segment is positively dependent on temperature
difference.  Segment G is the control segment, which is connected
to the junction particle between segments S and D. Temperature
$T_G$  will be used to control temperature $T_o$ (at the junction
O) so as to control the heat current from D to S. In analogy to
the MOSFET, in which the electronic current in the electrode G is
very small, we require here that the heat current $J_G$ through
segment G to be as small as possible, (otherwise it is hard to set
$T_G$ to a required value in experiment). Moreover the heat
resistance of segment G must be small enough in order to well
control the temperature $T_o$ by changing $T_G$ so that
$T_o\approx T_G$.

Notice that, in typical situations, the differential heat
resistance, $R_S=\left(\frac{\partial J_S}{\partial
T_o}\right)^{-1}_{T_S=\mbox{const}}$ in segment $S$, and
$R_D=-\left(\frac{\partial J_D}{\partial
T_o}\right)^{-1}_{T_D=\mbox{const}}$ in segment D,  are both
positive and therefore there exists only one value of $T_o$ for
which $J_S=J_D$ so that $J_G=0$, Since $J_S=J_D+J_G$, the
 ``{\it current amplification factor}",
$ \alpha=\left|\frac{\partial J_D}{\partial
J_G}\right|=\left|\frac{R_S}{R_S+R_D}\right| <1, $ namely in order
to make a change $\Delta J_D$, the control heat bath has to
provide a larger $\Delta J_G$. This means that the ``transistor''
can never work!

The key point of our transistor model is the ``{\it negative
differential heat resistance}'' as we observed in the diode
model\cite{diode}. It provides the possibility that when $T_o$
changes both $J_S$ and $J_D$ change simultaneously in the same
way. Therefore $J_S=J_D$ (or $J_s\approx J_D$) can be achieved for
several different values of $T_o$ or even in a wide region of
$T_o$ as shown in Figs.10 and 11. In this situation heat switch
and heat modulator/amplifier are possible. In the ideal, limiting
case of $R_S=-R_D$ which, in principle, can be obtained by
adjusting parameters, the transistor works perfectly.

\begin{figure}
\begin{center}
\includegraphics[width=7cm]{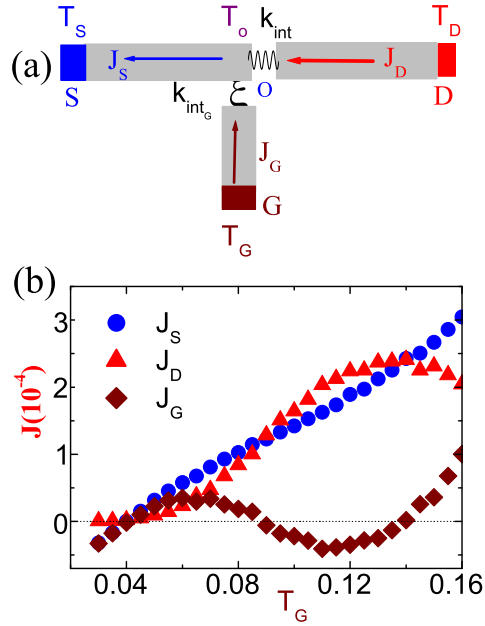}
\vspace{-.5cm}
 \caption{(a) Configuration of the
thermal transistor. (b) Heat current versus the control
temperature $T_G$. Parameters are: $T_D=0.2,, V_D=1.0, k_D=0.2,
k_{int}=0.05; T_S=0.04, V_S=5, k_S=0.2, V_G=5, k_G=1,
k_{int_G}=1$. Notice that both $J_S$ and $J_D$ inrease when the
temperature $T_G$ is increased. }
\end{center}
\end{figure}

\subsubsection{Thermal Switch}

We first demonstrate the ``\textit{switch}" function of our
transistor, namely we show that the system can act like a good
heat conductor or an insulator depending on the control
temperature. This is illustrated in Fig.10(b), where we plot $J_G,
J_S$, and $J_D$ versus $T_G$.  When $T_G$ increases from $0.03$ to
$0.135$, both $J_D$ and $J_S$ increase. In particular, at three
points: $T_G\approx 0.04, 0.09$ and $0.135, J_D=J_S$ thus $J_G$ is
exactly zero. These three points correspond to ``off", ``semi-on"
and ``on" states, at which $J_D$ is $2.4\times 10^{-6}, 1.2\times
10^{-4} $  and $2.3\times 10^{-4}$, respectively. The ratio of the
heat current at the ``on" state and that at the ``off" state is
about 100,  hence our model displays one important function -
switch - just like the function of a MOSFET used in a digital
circuit.

\begin{figure}
\begin{center}
\includegraphics[width=7cm]{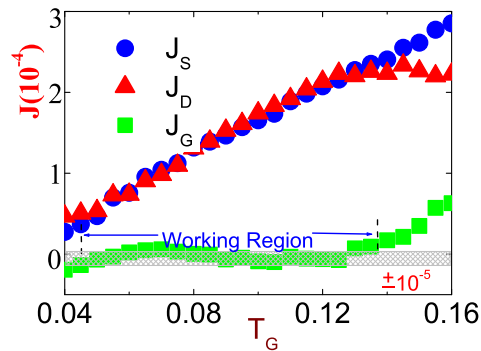}
\vspace{-.5cm} \caption{ Heat current versus the control
temperature $T_G$. Here: $T_D=0.2,  V_D=1.0, k_D=0.2, T_S=0.04,
V_S=5, k_S=0.2, k_{int}=0.05, V_G=5, k_G=1, k_{int_G}=0.1$. The
shadow region is the range of variation of $J_G$ in the
temperature interval $T_G\in(0.05, 0.135)$. }
\end{center}
\end{figure}
\subsubsection{Thermal modulator/amplifier}

As demonstrated above, the heat current from D to S can be
switched between different values. However, in many cases, like in
an analog circuit, we need to continuously adjust the current
$J_S$ and/or $J_D$ in a wide range by adjusting the control
temperature $T_G$. In Fig.11 we demonstrate this
``modulator/amplifier" function of our transistor. The basic
mechanism of such ``modulator/amplifier" is the same as that of
the ``switch" but we consider here different parameter values. It
is seen that in the temperature interval $T_G\in (0.05, 0.135)$,
the heat current through the segment G  remains very small
(($-10^{-5}\sim10^{-5}$), within the shadow strip in Fig.10, while
the heat currents $J_S$ and $J_G$ continuously increase from
$5\times 10^{-5}$ to $2\times 10^{-4}$.

\section{Conclusions and discussions}

In this paper, we have given a brief summary of our recent work on
heat conduction in one dimensional systems. We have shown that
strong chaos is sufficient but not strictly necessary for the
validity of the Fourier heat law. Indeed linear mixing can be
sufficient to induce a diffusive process which ensures normal heat
conductivity.

For systems with total momentum conservation one typically finds
anomalous conductivity, namely the thermal conductivity is
divergent with the system size. Anomalous conductivity has been
connected with anomalous diffusion via the very simple formula
(4).

Finally we have shown the possibility to build a thermal diode
which exhibits a very significant rectifying effect in a very wide
range of system parameters. Moreover, based on the phenomenon of
\textit{negative differential thermal resistance} observed in the
thermal diode, we have built a theoretical model for a thermal
transistor. The model displays two basic functions of a
transistor: switch and modulator/amplifier. Although at present it
is just a model we believe that, sooner or later, it can be
realized in a nanoscale system experiment.  After all the
Frenkel-Kontorova model used in our simulation is a very popular
model in condensed matter physics\cite{Fkreview1}.
\bigskip

\subsubsection{Acknowledgement}
BL is supported in part by Faculty Research Grant of National
University of Singapore  and the Temasek Young Investigator Award
of DSTA Singapore under Project Agreement POD0410553.  GC is
partially supported by EU Contract No. HPRN-CT-2000-0156 (QTRANS)
and by MURST (Prin 2003, Ordine e caos nei sistemi estesi non
lineari: strutture, stocasticita' debole e trasporto anomalo).

\end{document}